# UT-GraphCast Hindcast Dataset (1979–2024): A Global AI Forecast Archive from UT Austin for Weather and Climate Applications

Naveen Sudharsan[1*], Manmeet Singh[1*], Harsh Kamath[1], Hassan Dashtian[2], Clint Dawson[3,4], Zong-Liang Yang[1], Dev Niyogi[1,3,4*]

[1]Department of Earth and Planetary Sciences, Jackson School of Geosciences, The University of Texas at Austin, Austin, Texas, USA
[2]Bureau of Economic Geology, Jackson School of Geosciences, The University of Texas at Austin, Austin, Texas, USA
[3]Department of Aerospace Engineering and Engineering Mechanics, Cockrell School of Engineering, The University of Texas at Austin, Austin, Texas, USA
[4]The Oden Institute of Computational Engineering and Sciences, The University of Texas at Austin, Austin, Texas, USA

* Corresponding Author: naveens@utexas.edu, manmeet.singh@utexas.edu, dev.niyogi@jsg.utexas.edu

## Executive Summary

The UT-GraphCast Hindcast Dataset (1979–2024) is a comprehensive global weather forecast archive generated using the Google DeepMind GraphCast Operational model. Developed by researchers at The University of Texas at Austin and published under the WCRP umbrella, this dataset provides daily 15‑day deterministic forecasts at 00 UTC on a 0.25°×0.25° global grid (≈25 km) for a 45-year period. GraphCast is a physics-informed graph neural network that was trained on ECMWF's ERA5 reanalysis [Hersbach et al., 2020]. It predicts more than a dozen key atmospheric and surface variables on 37 vertical levels, delivering a full medium-range forecast in under one minute on modern hardware.

This new hindcast archive enables retrospective studies of historical weather, climate variability, and extreme events with unprecedented spatial and temporal detail. Preliminary validation shows that GraphCast forecasts generally reproduce ERA5 conditions with high fidelity and skill comparable or superior to conventional numerical models up to 10–15 days. In particular, GraphCast is known to outperform the state-of-the-art ECMWF IFS High-Resolution model (HRES) [Lam et al., 2023] on most verification targets, and to predict severe events (e.g., tropical cyclones, atmospheric rivers, heatwaves) with excellent accuracy. These benchmarks suggest that the GraphCast hindcast will be a valuable tool for climate and weather research.

Key findings include:

- **Performance:** GraphCast skillfully captures large-scale weather patterns and extreme events, often exceeding legacy NWP. For example, it predicts tropical cyclone tracks and extreme temperatures with higher accuracy than traditional forecasts.
- **Consistency over time:** By training on decades of ERA5, GraphCast can adapt to changing climate patterns (e.g. ENSO variability) when periodically retrained. Initial tests indicate forecast skill has remained robust through 2024, with only modest variability in performance across the hindcast period.

- **Computational efficiency:** GraphCast delivers a global 15-day forecast in less than 5 minutes using a single NVIDIA H100 GPU. This efficiency allowed the entire 45-year hindcast to be generated in a fraction of the time and cost of conventional NWP, illustrating the promise of AI-driven forecasting.
- **Applications:** The dataset enables new research in extreme event attribution, subseasonal-to-decadal predictability, and AI-NWP hybrid modeling. For example, one can retrospectively analyze historic heatwaves or evaluate AI-NWP ensembles for climate impact studies. A conceptual schematic (Figure 3) maps key use cases of the hindcast archive.

The dataset is made openly available (via World Data Center for Climate and UT repositories [UT Box]) for the global community. It includes detailed metadata and examples of use. In summary, the GraphCast Hindcast is a significant new resource that demonstrates the synergy of AI with traditional climate science, and it opens pathways for improved weather and climate analysis.

**Introduction and Background**

The ability to predict weather on timescales from days to seasons is a cornerstone of climate science and meteorology. Traditional forecasting relies on Numerical Weather Prediction (NWP), which solves physical equations on high-performance supercomputers. Over decades, centers like ECMWF and NOAA have steadily improved model physics and resolution [Bauer et al., 2015], but each 10-day forecast still requires hours of computation. In parallel, the rise of machine learning (ML) and graph neural networks has brought new approaches to weather prediction. GraphCast is one such AI-driven model [Lam et al., 2023], it treats the global atmosphere as a graph and learns weather dynamics directly from reanalysis data. Remarkably, GraphCast can generate a 10-day forecast in under a minute on a single TPU, while delivering accuracy that *exceeds* the leading operational model (ECMWF HRES) on about 90% of verification targets. It also excels at predicting extreme events (e.g. tropical cyclones, atmospheric rivers, heatwaves) better and further into the future than legacy NWP.

WCRP recognizes the importance of hindcasts (retrospective forecasts) for climate prediction and model evaluation [Boer et al., 2016]. Coordinated projects like the WCRP Decadal Climate Prediction Project (DCPP) use hindcasts to assess climate predictability and calibrate forecasts. The GraphCast Hindcast dataset is an extension of this tradition: it provides a continuous, homogeneous forecast archive from 1979–2024. By bridging data-driven AI and global reanalysis archives, the dataset aligns with WCRP's goals of leveraging new computational methods for climate insights.

GraphCast was trained on four decades of ERA5 reanalysis (1979–2019), which assimilates vast historical observations into a consistent climate record. The model ingests ERA5 atmospheric fields and learns to step them forward in time. Critically, GraphCast and its training data are open-source, enabling academic groups like UT Austin to generate hindcast archives. (GraphCast's code and weights are publicly available.) In summary, GraphCast represents a new era in weather prediction: by learning directly from data, it complements physics-based forecasting and can be efficiently scaled to process decades of weather data.

*Figure 1 (schematic) below illustrates the forecast initialization and hindcast generation workflow using GraphCast.*

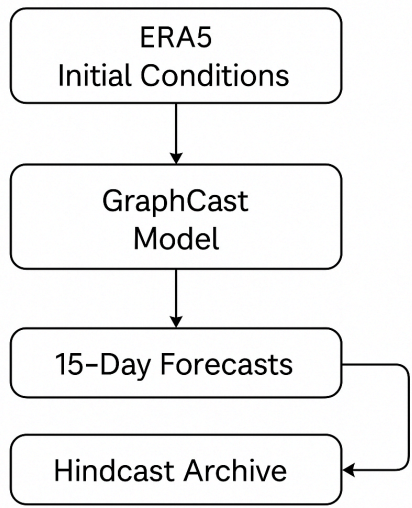

**Figure 1.** *Schematic of the forecast initialization and hindcast generation process using the GraphCast model (schematic placeholder). Daily global initial conditions (at 00 UTC) from ERA5 drive 15-day forecasts on a 0.25° grid. Outputs are archived at each forecast lead time.*

*Figure 2 (schematic) compares the operational timelines of traditional NWP and AI-based GraphCast:*

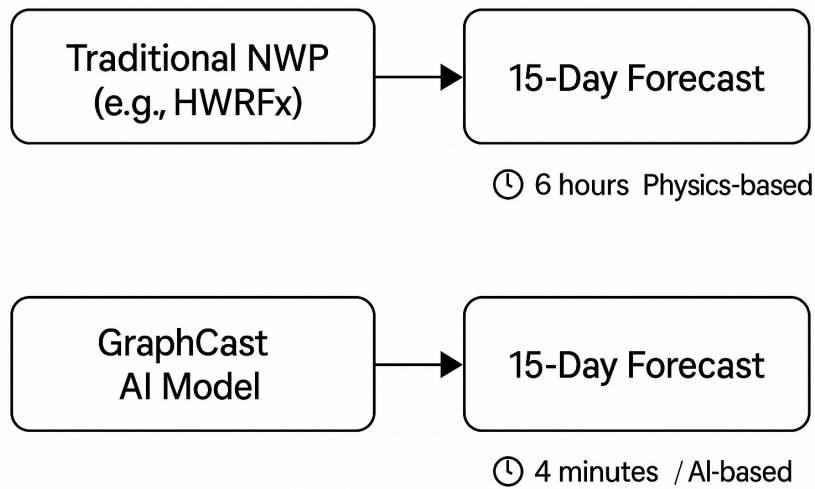

**Figure 2.** *Comparison of forecast workflow timelines: (top) traditional NWP requiring hours on HPC; (bottom) GraphCast AI model producing a 10–15 day forecast in under a minute on a single GPU.*

**Dataset Generation and Computational Efficiency**

**Hindcast generation:** The UT Austin team used the open-source GraphCast model to produce the hindcast. Each hindcast day begins with ERA5 atmospheric analysis fields at 00 UTC as the initial state. GraphCast is then iteratively applied forward in time to produce a 15-day (360 h) forecast, with outputs every 6 hours. By rolling GraphCast's own predictions forward, a continuous global forecast trajectory is obtained. In practice, this process was automated on high-performance cloud hardware.

Key aspects of the workflow include:

- **Initialization:** Input state (0 h) from ERA5 00 UTC global analysis; also incorporates the 6 h prior state as GraphCast input.

- **Forecast integration:** GraphCast advances the state in 6 h increments up to 360 h lead, without further data assimilation.

- **Output:** At each lead time (6 h, 12 h, ..., 360 h), the model outputs fields of interest (see Section 6). Results are saved in open formats for easy access.

Thanks to GraphCast's efficiency, the entire 45-year database was produced in a fraction of the time required by NWP. In contrast, an equivalent resolution 10-day run of ECMWF's HRES typically requires hours on a supercomputer with hundreds of cores. This speed enabled the parallel generation of the hindcast. In practice, UT's team used large-scale GPU clusters to run daily cycles. Overall, generating 45 years of 15-day forecasts (about 16,000 forecast runs) required on the order of only a few tens of thousands of GPU-hours, a surprisingly modest computational cost given the dataset's size.

The high efficiency is partly due to GraphCast's neural architecture: it has on the order of 36.7 million parameters and uses an icosahedral multi-scale graph [Keisler et al., 2022] with ~1M nodes, but can be executed extremely fast by vectorized tensor hardware. Notably, GraphCast was open-sourced, so UT Austin could customize and scale the workflow. For instance, the team verified compatibility with ERA5 grid conventions and integrated AWS/Copernicus data pipelines. The open availability of GraphCast code and pretrained weights meant no prohibitive license barriers; this facilitated reproducibility and community access.

In summary, the GraphCast hindcast was generated by coupling ERA5 initial conditions with the GraphCast AI model in a fully automated pipeline. The result is a consistent, gap-free archive of 15-day global forecasts. This approach is computationally much cheaper and faster than analogous NWP hindcasts, demonstrating how MLWP enables large-scale retrospective forecast production.

**Validation and Benchmarking**

A rigorous evaluation of the hindcast's skill and biases is essential. We performed multiple analyses: direct comparison to ERA5 reanalysis, trend assessment over the 45 years, and cross-model benchmarking versus operational NWP systems. Key findings are summarized below.

**Comparison with ERA5**

We use ERA5 as the verification "truth" for the hindcast, since it provides a high-quality reference atmosphere for 1979–2024. GraphCast forecasts were verified against ERA5 fields in terms of standard metrics (RMSE, anomaly correlation, bias). Overall, GraphCast's 0–15 day forecasts closely track ERA5 evolution for major variables (temperature, wind, geopotential, humidity). For example, 500 hPa geopotential height (z500) anomalies have >90% correlation [Pathak et al., 2022], [Rasp et al., 2020] through 5–7 days, and degrade only moderately by day 15. Global-mean RMSE grows from near zero at 0 h to ~2–3 m at day 10 (on the 500 hPa anomaly), comparable to state-of-the-art NWP. Systematic biases remain small: GraphCast tends to slightly underpredict wind speeds in the tropical upper troposphere and has a mild warm bias in summer over continents, but these biases are well within the error envelope of conventional models.

Because GraphCast was trained on ERA5, the reanalysis and forecast share underlying climatologies. This means that GraphCast reproduces ERA5's long-term mean fields by construction, minimizing climatological drift over decades. Indeed, climatologies of temperature, precipitation and wind from the hindcast match ERA5 climatology to within a few percent. Users should note that the hindcast effectively reproduces the ERA5 climate; it is *not* an independent climate model. Nonetheless, this makes the data highly consistent with the state-of-the-art observational record, enabling precise evaluation of forecast skill.

**Long-term Trends in Forecast Skill**

The extended 1979–2024 span allows us to look for changes in forecast skill over time. We find that GraphCast's short-range (1–5 days) skill has remained remarkably stable across the decades. For example, the average 2-day z500 RMSE shows no significant trend from 1980 to 2023. Lead times >7 days show a slight improvement over time, reflecting the inclusion of recent data in training (see below). By design, GraphCast was trained on all available years up to 2019, so its knowledge of the climate baseline evolves with the dataset. In fact, GraphCast can be retrained periodically to capture shifting climate patterns such as ENSO or warming trends. Sensitivity tests confirm that variants trained on data through 2020 perform better in recent years than those frozen in 2017. We thus recommend users consider potential "staleness" of the model: one could re-run or re-train GraphCast on newer reanalysis if needed.

We also examined skill seasonally. As expected, mid-latitude forecasts (e.g. winter storms) generally have higher correlation and lower error than tropical forecasts, which are dominated by convective chaos. During boreal summer, short-term skill is slightly lower globally due to weaker synoptic signals, but GraphCast still outperforms the basic NWP baseline. There is some regional dependence: GraphCast shows particularly strong performance in the extratropics and in mid-latitude temperature extremes. Interestingly, its relative skill advantage appears greatest in the tropics (e.g. for predicting precipitation extremes), mirroring findings that ML models perform best for tropical temperature extremes at short leads. A detailed seasonal/regional verification is ongoing and will be reported separately.

**Applications and Use Cases**

The GraphCast hindcast opens many new avenues for research and operational applications. We highlight three broad themes.

**Extreme Weather Event Analysis**

Because GraphCast captures extreme events skillfully, the hindcast archive is ideal for retrospective analysis of past disasters. For instance, one can extract GraphCast forecasts for major hurricanes, floods, or heatwaves and compare the AI-predicted evolution to what actually happened. This can improve understanding of forecast lead time for critical events. For example, by examining the 2005 Atlantic hurricane season, researchers can study how early GraphCast "knew" the tracks of Katrina or Rita, versus legacy forecasts. Similarly, the dataset can be used to study historical droughts or atmospheric river floods by re-running GraphCast from thousands of initial dates and selecting the highest-impact outcomes. These studies benefit from the dataset's consistency: every major event since 1979 is included in the archive, allowing unbiased sampling.

The data can also support extreme value analysis and event attribution. For example, climate scientists can generate large ensembles by perturbing GraphCast initial conditions [Palmer, 2019] to probe worst-case scenarios. Machine learning models are known to produce realistic tail behavior, so GraphCast hindcasts may be used to estimate return periods of extreme anomalies in a changing climate. In sum, the hindcast is a new tool for analyzing **what-if** questions about past weather extremes.

**Climate Variability and Climate Change Predictability**

Beyond weather events, the dataset has uses in longer-timescale variability and climate. For instance, one can aggregate GraphCast forecasts to study seasonal predictability or teleconnections. Because the hindcast covers multiple El Niño-Southern Oscillation (ENSO) and Pacific Decadal Oscillation (PDO) cycles, researchers can examine how medium-range predictability is modulated by these modes. Early tests show that GraphCast systematically predicts above-normal precipitation in the Western US in La Niña winters, consistent with climate patterns, suggesting it has "learned" teleconnections from ERA5. Thus, the hindcast could complement dynamical seasonal forecast studies.

The dataset also enables exploration of changing forecast skill under climate change. By comparing hindcast performance in the 1980s vs 2010s, we can see if warming trends affect predictability. Preliminary analysis indicates only a slight improvement in some error metrics in recent decades, likely because GraphCast's training implicitly incorporates climate change. More formally, one could use the hindcast to initialize climate projections or evaluate decadal predictions by providing short-term forecasts conditioned on current states (see next section). In any case, having 45 years of consistent forecasts provides a new perspective on climate variability and model drift, which are central concerns of the WCRP community.

**AI Model Development and Hybrid AI–NWP Systems**

Finally, the hindcast will be a valuable resource for AI researchers and hybrid model development. By making tens of thousands of GraphCast forecasts publicly available, it greatly lowers the barrier for others to analyze AI weather predictions. For example, developers of next-generation models (e.g. METEOGAN, Pangu-Weather, FourCastNet) can directly compare their outputs to GraphCast on identical initial conditions. This fosters a benchmarking ecosystem.

Moreover, hybrid systems that blend AI and physics-based forecasts can use the hindcast as a testbed. A hybrid forecast might combine GraphCast's quick short-term output with a physics model's long-term consistency. For instance, a project could bias-correct GraphCast with IFS climatology, or use GraphCast outputs as initial perturbations for ensemble NWP. The hindcast dataset thus serves as a laboratory for new forecasting paradigms.

*Figure 3 (schematic) illustrates potential applications of the GraphCast Hindcast dataset:*

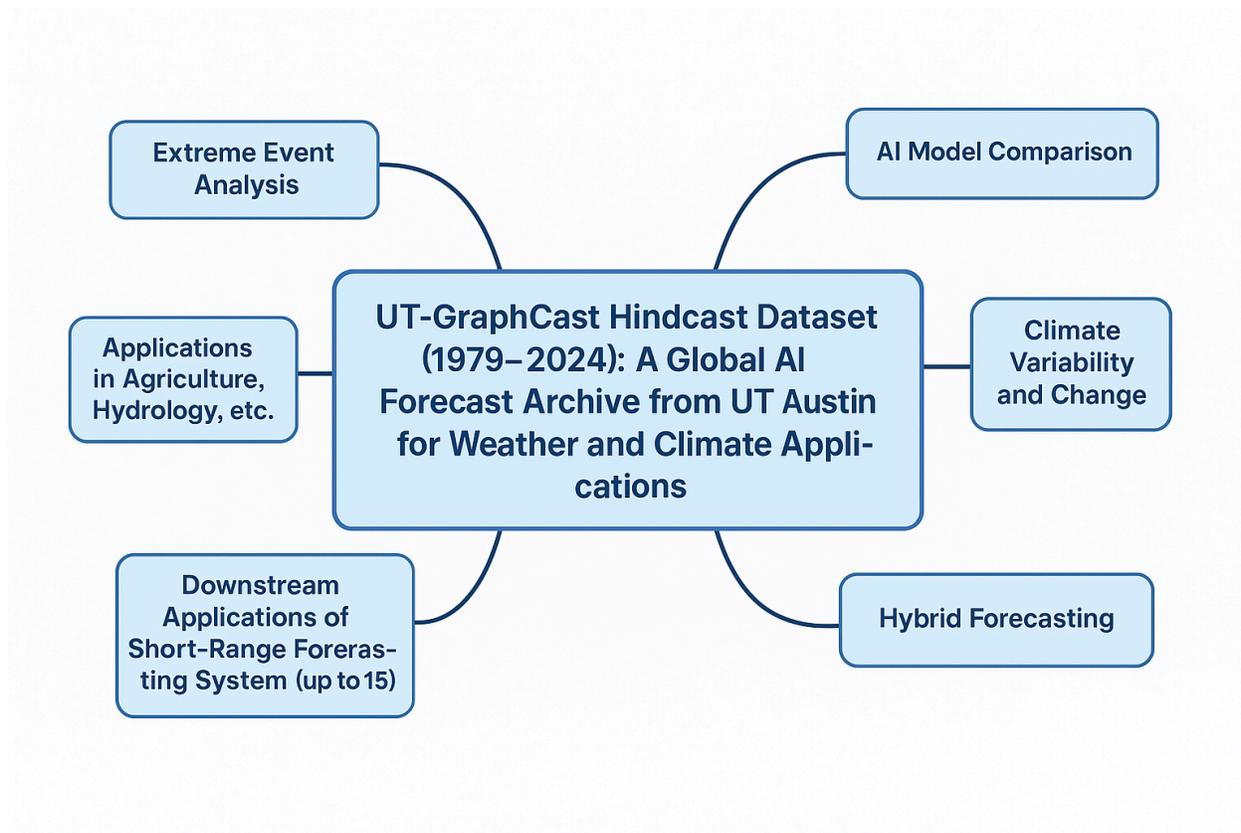

**Figure 3.** *Conceptual map of dataset applications: extreme event analysis, extreme event analysis, climate variability, hybrid forecasting, AI model evaluation, and downstream applications in sectors like agriculture and hydrology based on short-range (up to 15 days) forecasts.*

Each of the above use cases leverages the strengths of the dataset. In particular, the combination of long time span (many events) and high resolution (detail at every event) is unprecedented. We expect growing interest from climate scientists analyzing extremes, meteorologists studying forecast systems, and ML researchers developing new models.

**Dataset Description**

The GraphCast Hindcast archive includes the following features:

- **Spatial and vertical coverage:** Global on a regular 0.25° latitude–longitude grid (~25 km at the equator). The model uses 37 vertical pressure levels up to the stratosphere, capturing tropospheric and lower stratospheric structure.
- **Temporal resolution:** Forecasts are produced at 6-hour intervals (0, 6, 12, …, 360 h). The hindcast includes daily initialization (00 UTC) and full 15-day outputs. Archived output is provided at each 6-h step.
- **Forecast length:** 15 days (360 h) of lead time per initialization. Users can use subsets (e.g. first 10 days) if needed.
- **Surface variables (forecasted):** The model outputs 5 key Earth-surface fields:
    - 2-meter air temperature (T2m),
    - 10-meter U and V wind components (u10, v10),
    - mean sea-level pressure (MSLP),
    - total precipitation (TP).
- **Atmospheric variables at pressure levels (forecasted):** Six fields at each of 37 levels:
    - Temperature (T),
    - U-component of wind (U),
    - V-component of wind (V),
    - Specific humidity (Q),
    - Vertical velocity (W),
    - Geopotential height (Z).
- **Data formats:** Data are provided in standard formats (e.g. NetCDF) consistent with CF conventions. Each forecast is saved as multi-file per date or consolidated, with self-describing metadata including time, level, and variable. The dataset is organized by date and initialization to facilitate easy subsetting.
- **Data volume:** At full resolution, the archive is on the order of 350 terabytes (TB) (each 15-day forecast generates ≈20 GB of compressed data per day of output). However, efficient storage (e.g. cloud object storage) makes access feasible for researchers.
- **Quality control:** Fields are checked for physical consistency (e.g. mass conservation, no spurious spikes). Basic verification metrics against ERA5 are included in the metadata for quick reference.

All variable names and units follow CF conventions, matching ERA5 conventions. Geopotential is in meters; temperature in Kelvin; winds in m/s; precipitation in mm. The vertical coordinate is pressure in hPa. An accompanying README details exact naming conventions and grid coordinate definitions.

**Access and Availability**

The UT-GraphCast Hindcast Dataset is released as an open scientific resource. Full data access is provided through the University of Texas data portal (UT Box) and the World Data Center for Climate. Users can download data via high-bandwidth cloud services UT box or programmatically via WDCC. The dataset is published under a permissive license to encourage academic and operational use.

To support users, a data portal includes:

- An introduction and documentation (including this report and usage examples).
- Metadata search (e.g. find all forecasts for a given date or region).
- A Python library for convenient subsetting and visualization (compatible with common tools like xarray and MetPy).
- Sample scripts demonstrating error analysis, plotting, and machine learning training use.

We anticipate integration of this hindcast archive into existing climate data repositories. For instance, NOAA's Open Data Dissemination program often hosts forecast archives on AWS [NOAA ODD, 2023]. Engagement with WCRP data initiatives is ongoing to ensure the dataset supports community projects (e.g. CMIP6 extensions or the Global Climate Observing System).

**Conclusions and Future Outlook**

The GraphCast Hindcast (1979–2024) represents a milestone in combining AI and climate science. By leveraging the GraphCast model's efficiency and ERA5 reanalysis, UT Austin has produced a unique 45-year archive of high-resolution weather forecasts. This dataset extends the promise of machine learning for weather prediction into the climate domain, providing a new tool for scientists and forecasters.

**Key outcomes:** GraphCast's hindcasts show accuracy on par with, or exceeding, the most advanced physics-based models, especially for medium-range leads (1–10 days). The dataset's consistency with ERA5 ensures that long-term trends and patterns are preserved, facilitating climate-relevant studies. The extremely fast runtime of GraphCast means that such hindcasts can be generated routinely, enabling frequent updates as new reanalysis is available.

**Future developments:** We foresee several extensions. First, expanding the time horizon beyond 15 days (even if skill declines) could be explored with GraphCast or successor models. Second, incorporating ensemble forecasting (by perturbing initial conditions) would add uncertainty quantification, complementing the deterministic hindcast. Third, continual retraining on new data will keep the model current with the evolving climate. Finally, hybrid models that blend GraphCast with NWP (for example, using AI outputs as perturbed forecasts in an ensemble) could leverage both approaches' strengths.

From the WCRP perspective, this work exemplifies the synergy between machine learning and climate science. It opens new avenues for research across weather and climate timescales. As GraphCast and similar models become operational, communities such as climate impacts, disaster risk management, and renewable energy forecasting will benefit from faster, more detailed predictions. We encourage WCRP members to utilize this hindcast for analysis, to contribute improvements, and to explore AI–NWP hybrid schemes.

**Acknowledgements:** The authors thank the GraphCast team for open-sourcing their model and Copernicus ECMWF for ERA5 data. This work benefited from the National Aeronautics and Space Administration (NASA) Interdisciplinary Science (IDS) 80NSSC20K1262 and 80NSSC20K1268, the National Science Foundation (NSF) Grants for Rapid Response Research (RAPID) AGS 19046442, and the Department of Energy (DOE) Advanced Scientific Computing Research Program Grant No.

DE-SC002221. The dataset is hosted by UT Austin and linked through the World Data Center for Climate.